\documentclass{ptephy_v1}

\preprintnumber{XXXX-XXXX} 

\newcommand{\susy}{{\bf Q}}

\begin{document}

\title{Localization principle in SUSY gauge theories}


\author{Kazuo Hosomichi}
\affil{Department of Physics, National Taiwan University, Taipei 10617, Taiwan \email{hosomiti@phys.ntu.edu.tw}}





\begin{abstract}%
Localization principle is a powerful analytic tool in supersymmetric gauge
theories which enables one to perform supersymmetric path integrals
explicitly. Many important formulae have been obtained, and they led to a
major breakthrough in the understanding of gauge theories at
strong coupling as well as the dynamics of branes in M-theory.
Some of those results are reviewed, focusing especially on Pestun's
solution to four-dimensional ${\cal N}=2$ supersymmetric gauge theories
on $S^4$ and the subsequent developments on three or four-dimensional
 gauge theories on spheres.
\end{abstract}

\subjectindex{xxxx, xxx}

\maketitle

\section{Introduction}\label{sec:intro}

Localization is a powerful mathematical principle that sometimes allows
us to reduce the difficulty of integrals over complicated
spaces. If a continuous symmetry acts on the space, one can express certain
integrals over that space as sums of contributions from fixed points,
that is the points which are invariant under the symmetry. It has been
applied to the problems in supersymmetric gauge theories in different
ways, and led to a number of useful formulae that can plobe the strong
coupling dynamics of gauge theories.

\paragraph{What is localization principle}

Let us explain the basic idea of localization principle, quoting an illustrative
example of the volume of sphere from \cite{Witten:1992xu}. We use
the standard polar coordinates $\theta,\phi$ on the sphere $S^2$, in
terms of which the simplectic volume form is given by
$\omega=\sin\theta d\theta d\phi$. Using the
rotational symmetry generated by the vector field $v=\partial_\phi$, one
can think of a deformation of the ordinary derivative $d$ into an
equivariant derivative $\susy\equiv d-\epsilon i_v$, and accordingly
deform the ordinary closed forms into the differential forms annihilated
by $\susy$, called equivariantly closed forms. The volume form $\omega$
is then modified into $\omega+\epsilon H$, where
$H=\cos\theta$ is the Hamiltonian function for the isometry $v$. The
symplectic volume of $S^2$ then receives the following modification,
\begin{equation}
 4\pi = \int\omega=\int e^\omega\quad\Longrightarrow\quad
 \int e^{\omega+\epsilon H}
 = \frac{2\pi}\epsilon(e^{\epsilon}-e^{-\epsilon})\,.
\label{eq:S2eq}
\end{equation}
Interestingly, in the rightmost expression the two terms can be
interpreted as contributions from two fixed points, namely the
north and the south poles. Indeed, one can ``approximate'' the
contribution from the north pole $\theta=0$ by a Gaussian integral over
local Cartesian coordinate $x,y$,
\begin{equation}
 \int dxdy e^{\epsilon\{1-\frac12(x^2+y^2)\}}=\frac{2\pi e^\epsilon}
 \epsilon\,,
\label{eq:S2NP}
\end{equation}
where $e^\epsilon$ is the classical value of $e^{\epsilon H}$ at the
north pole and $2\pi/\epsilon$ is the result of Gaussian integration
over $x,y$.
The same approximation at the south pole and suitable Wick rotation of
the integration contour can explain the other term.
This example is the simplest application of localization principle or
Duistermaat-Heckman's formula in mathematics.

In string theory or quantum field theories, complicated spaces often
arise as moduli space of solutions to some field equations such as BPS
conditions in supersymemtric models. The integrals over such
moduli spaces often provide a useful low-energy approximation to the
original path integral. Actually, in some supersymmetric
theories one can deform the theory in a suitable manner so that
the moduli space approximation becomes exact. An example of such
deformations is the topological twist of four-dimensional ${\cal N}=2$
supersymmetric gauge theories, which was invented for studying
the cohomology of instanton moduli spaces within the framework of quantum
field theory \cite{Witten:1988ze}. Another example is the topological
A-twist of 2D supersymmetric sigma models and the corresponding
topological string, which involve integrals over moduli space of
holomorphic maps from Riemann surfaces to Calabi-Yau manifolds
\cite{Witten:1988xj}.

In the above examples, the reduction from infinite-dimensional path
integrals to finite dimensional integrals makes use of the idea of
localization based on a fermionic symmetry (supersymmetry) $\susy$. The
supersymmetry means the action functional $S$ is invariant under
$\susy$, and also the path integral measure is such that the expectation
values of $\susy$-exact observables all vanish.
\begin{equation}
 \langle \susy(\cdots)\rangle ~=~ \int e^{-S}\susy(\cdots)
 ~=~ \int \susy(e^{-S}\cdots)=0.
\end{equation}
These imply that the values of supersymmetric observables do not change
under deformations of the theory of the form $S\to S+t\susy{\cal V}$ for
arbitrary parameter $t$ and fermionic functional ${\cal V}$ such that
$\susy^2{\cal V}=0$. The supersymmetric path integrals thus localize to
{\it saddle points} characterized by the BPS-like condition
\begin{equation}
 \susy\Psi=0\quad\text{for all the fermions}~\Psi.
\end{equation}
To evaluate the contribition from each saddle point, one only needs to
path integrate over fluctuations with Gaussian approximation, keeping
only terms in $\susy{\cal V}$ up to the second order in the fluctuations
as was done in (\ref{eq:S2NP}). This gives an exact answer
because the supersymmetric observables are $t$-independent.

Thus in supersymmetric theories, localization is applied for two
different purposes. One is the reduction from infinite-dimensional path
integral to a finite dimensional integral over moduli spaces (called
{\it SUSY localization} in this article), and the other is the
simplification of integrals over complicated moduli spaces using
symmetry ({\it equivariant localization}). The underlying principle is
the same: in particular they are both characterized by a fermionic
operator $\susy$ which squares to a bosonic symmetry of the system.

\paragraph{Localization in SUSY gauge theories}

The two kinds of localization both played
important role in \cite{Nekrasov:2002qd} where Nekrasov proposed the
topologically twisted gauge theory on Omega background
$\mathbb{R}^4_{\epsilon_1,\epsilon_2}$. In this theory, the
supercharge $\susy$ squares into a spacetime rotation plus a constant gauge
rotation,
\begin{equation}
 \susy^2 = \epsilon_1J_{12}+\epsilon_2J_{34}+\text{gauge}(a).
\end{equation}
where $a$ is the expectation value of the scalar field in vector multiplet
which parametrizes the Coulomb branch moduli space. The path integral of
this theory defines the so-called Nekrasov's instanton partition
function, which is the generating function for equivariant integrals
over instanton moduli spaces. The parameters $\epsilon_1,\epsilon_2,a$
play the same role as that of $\epsilon$ in (\ref{eq:S2eq}), and
simplifies the integrals over instanton moduli spaces to combinatoric
sums. Nekrasov's partition function is known to contain the information
on the low energy effective prepotential; in fact there is an extensive
study showing it encodes even richer information on the mathematical
structure underlying 4D ${\cal N}=2$ supersymmetric gauge theories. See
\cite{Itoyama:PTEP} for a review on this field.

In 2007, the idea of SUSY localization was first applied to gauge
theories which are not topological field theories. In
\cite{Pestun:2007rz} Pestun used the localization principle to obtain an
exact formula for supersymmetric observables in 4D ${\cal N}=2$ SUSY
gauge theories on the sphere $S^4$. He showed that the infinite
dimensional path integral can be reduced to a finite-dimensional
integral over Lie algebra, and using the result he gave an analytic proof
of the long standing conjecture about the Wilson loops in ${\cal N}=4$
super Yang-Mills theories \cite{Erickson:2000af,Drukker:2000rr}. In
2009, another exact formulae was found for 3D superconformal
Chern-Simons matter theories by Kapustin, Willett and Yaakov
\cite{Kapustin:2009kz}. Together with the application of localization to
the 3D superconformal index by Seok Kim \cite{Kim:2009wb}, these work
brought the power of localization to the attention of many physicists.

\vskip3mm

This article is a brief review of the pioneering work
\cite{Pestun:2007rz} and \cite{Kapustin:2009kz} and the subsequent
developments in supersymmetric gauge theories based on localization
principle. In the first part we will focus mostly on theories in 4 and 3
dimensions and the developments around exact partition functions on
sphere. In the latter part we will discuss interesting developments
regarding supersymmetric deformation of the round sphere called squasings.

\paragraph{Remark}

SUSY localization reduces the path integral
to an integral over the space of saddle points, and allows us to
treat the fluctuations around saddle points by Gaussian approximation.
The Gaussian integral in field theory gives rise to determinants of
Laplace or Dirac operators, which are usually defined as infinite
products over eigenvalues. In the following we will see many formulae for the
determinants. On the face of it those infinite product formulae do not
make sense or they are simply diverging, but they do make sense by a
suitable regularization. Let us not worry too much about the
regularization issue, instead recall that the same kind of infinite
product arises even for the path-integral evaluation of partition
function for a single harmonic oscillator.
\begin{equation}
\int{\cal D}q(t)\exp\left[-\frac1\hbar\int_0^\beta
 dt\left(\frac12m\dot q^2+\frac12m\omega^2q^2\right)\right]
 ~=~
 \text{(const)}\cdot
 \prod_{n\in\mathbb{Z}}\frac1{\beta\omega+2\pi i n}.
\end{equation}
The infinite product is understood as the result of path integration
over Fourier modes of the periodic variable $q(t)\sim q(t+\beta)$.
It needs an appropriate regularization so as to reproduce the
desired result $1/2\sinh(\beta\omega/2)$.

\section{4D ${\cal N}=2$ gauge theories}

For 4D ${\cal N}=2$ supersymmetric gauge theories, exact partition
function was obtained for topologically twisted theories on Omega
background in \cite{Nekrasov:2002qd}. Based on this result, Pestun
\cite{Pestun:2007rz} obtained the closed formula for supersymmetric
observables on $S^4$. A little later there was a development in the construction
and classification of superconformal theories based on the picture of
wrapped M5-branes, which led to a new understanding of the relation
between ${\cal N}=2$ gauge theories and the geometry of Riemann surface
\cite{Gaiotto:2009we}. These development also led to a discovery of a
surprising relation between observables in 4D gauge theories and 2D
conformal field theories \cite{Alday:2009aq,Wyllard:2009hg}. 

\paragraph{Exact solution on $S^4$}

Let us begin by reviewing the exact results for the theories on $S^4$.
In \cite{Pestun:2007rz}, the theories was constructed by using the
conformal map from flat $\mathbb{R}^4$. The supersymemtry is
characterized by conformal Killing spinors
$\xi_{\alpha A},\bar\xi^{\dot\alpha}_A$ satisfying
\begin{eqnarray}
 D_m\xi_A \equiv
 \left(\partial_m+\frac14\Omega_m^{ab}\sigma_{ab}\right)\xi_A
 = -i\sigma_m\bar\xi'_A,&~&
 D_m\bar\xi'_A = -\frac{i}{4\ell^2}\bar\sigma_m\xi_A,
 \nonumber \\
 D_m\bar\xi_A \equiv
 \left(\partial_m+\frac14\Omega_m^{ab}\bar\sigma_{ab}\right)\bar\xi_A
 = -i\bar\sigma_m\xi'_A,&~&
 D_m\xi'_A = -\frac{i}{4\ell^2}\sigma_m\bar\xi_A.
\label{eq:CKS4}
\end{eqnarray}
The indices $\alpha,\dot\alpha$ (usually suppressed) represent they are
spinors under four-dimensional rotation group, whereas the index $A$ is
for doublets under $SU(2)$ R-symmetry. See \cite{Hama:2012bg} for our
convention of spinor calculus here. In theories with rigid
supersymmetry on curved spaces such as spheres, these Killing spinors
appear in SUSY transformation rule in place of constant spinor
parameters. Unlike the supersymmetry parameters for theories on flat
$\mathbb{R}^4$ they are in general not constant. But they take a fixed form
once the diffeomorphism and other local gauge invariance are fixed.

${\cal N}=2$ theory has two supermultiplets. Vector multiplet consists
of a vector $A_m$, a complex scalar $\phi$, gauginos
$\lambda_{\alpha A},\bar\lambda_{\dot\alpha A}$ and auxiliary scalar
fields. Hypermultiplet consists of an $SU(2)_\text{R}$ doublet scalar
$q_A$, fermions $\psi_\alpha,\bar\psi_{\dot\alpha}$ and auxiliary fields.
Once the gauge group $G$ and representation $R$ for hypermultiplet are
specified, one can construct supersymmetric Laglangian for vector multiplet,
\begin{equation}
 S_{\text{YM}} = \int d^4x\sqrt{g}
 \text{Tr}\left(\frac1{2g^2}F_{mn}F^{mn}
 +\frac{i\theta}{32\pi^2}\varepsilon^{klmn}F_{kl}F_{mn}+\cdots\right),
\end{equation}
and the kinetic Lagrangian for hypermultiplets coupled to vector multiplets.
One can also include the hypermultiplet mass term (or other SUSY
invariant called FI term which we will not discuss here) in
the action. The partition function will then be a
function of gauge coupling $\tau=\frac{\theta}{2\pi}+\frac{4\pi i}{g^2}$,
the matter mass $m$ and the radius $\ell$ of the sphere.

The existence of SUSY theories on spheres was known and even used in the
study of superconformal indices or construction of superstring
worldsheet theories. But the notion of conformal Killing spinors and the
fully explicit construction of supersymmetric gauge theories on the sphere
looked new and rather surprising.

To apply SUSY localization, one first chooses a specific Killing spinor
$\xi_A,\bar\xi_A$. For generic choice there are two special points on
$S^4$, the north and south poles, characterized respectively by
$\xi_A=0$ and $\bar\xi_A=0$. If the $S^4$ is defined by
\begin{equation}
 x_0^2+x_1^2+x_2^2+x_3^2+x_4^2=\ell^2,
\end{equation}
then one can put the north pole at $x_0=\ell$ and the south pole at
$x_0=-\ell$ using conformal symmetry. The square of the corresponding
supersymmetry yields the sum of rotations about $(x_1,x_2)$-plane and
$(x_3,x_4)$-plane with equal coefficients. In particular, near the two
poles the supersymmetry is approximately that of topologically (anti-)twisted
theory with Omega deformation $\epsilon_1=\epsilon_2=\ell^{-1}$.

The supersymmetric saddle points are given by the constant value $a$
of the scalar in vector muliplet, and the hypermultiplet fields have to
be all zero. Gauge field is also required to be zero at generic points
on $S^4$ up to gauge choice, but it can take point-like instanton or
anti-instanton configuations at the north or south poles.
The (anti-)instantons give rise to Nekrasov's partition functions from
each pole. Thus the full partition function takes the form
\begin{equation}
 Z = \int d{a}e^{-S_\text{YM}(\tau;a)}Z_\text{1-loop}(a,m)
 Z_\text{Nek}(q;a,m,\epsilon_1,\epsilon_2)
 Z_\text{Nek}(\bar q;a,m,\epsilon_1,\epsilon_2)\,.
\end{equation}
Here the integral is over Cartan subalgebra of $G$, $m$ is the matter
mass and $q=e^{2\pi i\tau}$ becomes the instanton counting parameter in
the Nekrasov's partition function.
The classical action and one-loop determinant are given by
\begin{equation}
 e^{-S_\text{YM}} = (q\bar q)^{\frac12\text{Tr}(a^2)},\quad
 Z_\text{1-loop} =
  \frac{\prod_{\alpha\in\Delta}\Upsilon(ia\cdot\alpha)}
       {\prod_{w\in R}\Upsilon(1+ia\cdot w+im)}\,,
\label{eq:Z1l4d}
\end{equation}
where $\alpha$ runs over the root of $G$ and $w$ is the weight of the
representation $R$. The function $\Upsilon(x)$ here is
defined as an infinite product
\begin{equation}
 \Upsilon(x)= (\text{const})\cdot\prod_{n\ge1}
 (x-1+n)^n(1-x+n)^n\,.
\end{equation}

As reviewed in the introduction, the one-loop determinant can be
evaluated by choosing a suitable $\susy$-exact deformation of the action
$\susy{\cal V}$, approximating it by a quadratic functional in
fluctuations and evaluating the Gaussian integral. However, the standard
choice of $\susy{\cal V}$ for this problem does not lead to quadratic
functionals which respect $SO(5)$ rotation invariance of $S^4$, so the
direct evaluation of the determinant is very complicated. An elegant
solution is to translate the problem into
that of the index of (transversally elliptic) differential operators,
which essentially evaluates the trace of $e^{-it\susy^2}$ on some
reduced Hilbert spaces. If one uses this idea, there is actually no need
to explicitly work out the spectrum of any Laplace or Dirac operators. A
detailed explanation of how to compute the indices for transversally
elliptic differential operators was given in \cite{Pestun:2007rz}
including subtle issues of regularizations. Though mathematically quite
involved, the use of index theorem has become essential in studying SUSY
gauge theories, especially in higher dimensions.

One of the main purposes to solve the SUSY gauge theories on $S^4$
was to give an analytic proof of the conjecture
\cite{Erickson:2000af,Drukker:2000rr} that circular Wilson loops in
${\cal N}=4$ super Yang-Mills theory is given by a Gaussian matrix
integral. To show this, one chooses the hypermultiplet to be in the adjoint
representation of $G$ and apply the result of localization to the
so-called ${\cal N}=2^\ast$ theory.
When the mass for the hypermultiplet is turned off, then the one-loop
determinant becomes nothing but the Vandermonde determinant. The
Nekrasov's partition function also becomes trivial $Z_\text{Nek}=1$.
Thus one can explicitly see that the path integral reduces to just the
Gaussian matrix integral over $a$.

\paragraph{AGT relation}

In 2009 there was a series of breakthrough in 4D ${\cal N}=2$
supersymmetric gauge theories. Gaiotto proposed the construction of
families of superconformal field theories of {\it class S} based on
the picture of multiple M5-branes wrapped on punctured Riemann surfaces
\cite{Gaiotto:2009we}. Interestingly, for these class models the
marginal gauge couplings can be identified with the complex structure
moduli of the Riemann surface wrapped by the M5-branes. This led to a
geometric interpretation of the strong-weak coupling dualities in gauge
theories.

A little later, Alday, Gaiotto and Tachikawa found a surprising
correspondence between a family of gauge theories of class S and two-dimensional
Liouville CFT \cite{Alday:2009aq}. They studied the theories
describing two M5-branes wrapped on Riemann surface $\Sigma$ with $n$
punctures. The $S^4$ partition function and the Nekrasov's partition
function of the resulting theory $T_{\Sigma}$ were then compared with
the $n$-point correlation function of Liouville theory on $\Sigma$ and
its holomorphic building blocks called conformal blocks, and they were
shown to agree precisely. Similar correspondence was found between
class-S theorys of higher rank and Toda conformal field theories by
\cite{Wyllard:2009hg}. See a review \cite{Tachikawa:PTEP} for more
detail on this correspondence.

Toda theories and the 6D theories on multiple M5-branes
both obey ADE classification. The theory on two M5-branes and
Liouville theory are both labeled by $A_1$, the
simplest entry in this classification. Let us summarize here
the essential facts in Liouville theory and then try to describe how an
expert in Liouville theory would have understood the AGT relation when
it was first proposed.

\paragraph{Liouville theory revisited}

Liouville theory is a
theory of a massless real scalar field $\phi$ with exponential potential
$e^{2b\phi}$, where $b$ is called Liouville coupling. Though
interacting, it is known to be a conformal field theory of central charge
\begin{equation}
 c=1+6Q^2,\quad Q=b+\frac1b.
\end{equation}
Another remarkable feature of Liouville theory is the self-duality: the
theories with couplings $b$ and $1/b$ are known to be equivalent.
Thanks to conformal symmetry, correlation functions of arbitrary set of
local operators on general Riemann surfaces can in principle be
constructed algebraically from the two and three-point functions of
primary operators on sphere \cite{Belavin:1984vu}.
The three-point function of primary operators
$V_\alpha\equiv \text{const}\cdot e^{2\alpha\phi}$ in Liouville theory
\begin{equation}
 \langle V_{\alpha_3}(\infty)V_{\alpha_2}(1)V_{\alpha_1}(0)\rangle
 = C^{(3)}_{\alpha_1,\alpha_2,\alpha_3},
\end{equation}
was obtained in \cite{Dorn:1994xn} and \cite{Zamolodchikov:1995aa}.

\begin{figure}[ht]
\begin{center}
\includegraphics[width=9cm]{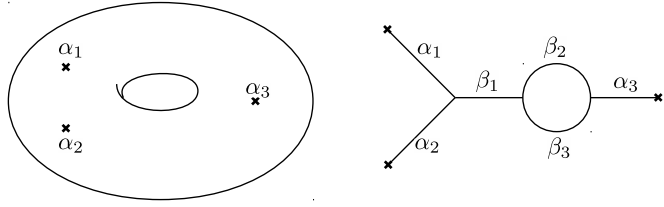}
\end{center}
\caption{Torus three-point function and its Moore-Seiberg graph}
\label{fig:msgraph}
\end{figure}

Conformal blocks are the basic building blocks in the construction of
correlators. In general, the dependence of correlation functions of 2D
CFT on the moduli $\tau_i$ of punctured Riemann surface (the shape of
the surface as well as the position of the insertions) is determined by
the conformal Ward identity. They consists of a set of holomorphic
differential equations in $\tau_i$ and the similar set for
$\bar\tau_i$. Conformal blocks are the solutions to the set of
holomorphic differential equations. There are different choices for the
basis of conformal blocks corresponding to different channels in which
to express correlators. For example, the diagram on the right of Figure
\ref{fig:msgraph}, called Moore-Seiberg graph, expresses the torus
three-point function
$\langle V_{\alpha_1}V_{\alpha_2}V_{\alpha_3}\rangle_{T^2}$ in a
particular channel, in which $\alpha_a$ are external Liouville momenta
and $\beta_a$ the momenta along the internal lines. The conformal blocks
${\cal F}$ in this channel are functions of $\alpha_a, \beta_a$ as well as
$\tau_i$. The correlation function can then be expressed as
\begin{equation}
\langle V_{\alpha_1}V_{\alpha_2}V_{\alpha_3}\rangle_{T^2}(\tau_i,\bar\tau_i)
 ~=~ \int d^3\beta\,C^{(3)}_{\alpha_1,\alpha_2,\beta_1}
 C^{(3)}_{\beta_1,\beta_2,\beta_3}C^{(3)}_{\beta_2,\beta_3,\alpha_3}
 |{\cal F}_{\vec\alpha,\vec\beta}(\tau_i)|^2\,.
\label{eq:corr}
\end{equation}

\noindent
Under AGT relation, the conformal blocks ${\cal F}$ are identified with
Nekrasov's partition function, and the product of $C^{(3)}$ with
one-loop determinants. The momenta $\vec\alpha$ and $\vec\beta$
correspond to masses $m$ and Coulomb branch parameters $a$. In particular, each
internal line in the Moore-Seiberg graph corresponds to an $SU(2)$
vector multiplet. With all these identifications understood, the formula
(\ref{eq:corr}) looks like an $S^4$ partition function \cite{Alday:2009aq}.

Let us now look into the correspondence in more detail.
First, it was proposed in \cite{Alday:2009aq} that the parameters
$\epsilon_1,\epsilon_2$ of Omega deformation is related to the Liouville
coupling $b$ as
\begin{equation}
 \epsilon_1:\epsilon_2 = b:\frac1b.
\end{equation}
This implies that the correspondence between Nekrasov's partition
functions and conformal blocks is for general Liouville central charge,
but the $S^4$ partition function should correspond to Liouville correlators
at a special (self-dual) value of Liouville coupling $b=1$, since the
Omega background with $\epsilon_1=\epsilon_2$ showed up near the poles.
A natural question, as was already raised in \cite{Alday:2009aq},
would have been what kind of deformation of $S^4$ would give the CFT
correlators at $b\ne 1$. That led to the idea of {\it squashing}.

Second, the one-loop determinant $Z_\text{1-loop}$ in the $S^4$
partition function was identified with the product of the Liouville
three-point function $C^{(3)}_{\alpha_1,\alpha_2,\alpha_3}$. The analytic
property of $C^{(3)}$ can be determined from the following physical
requirements of Liouville theory.
\begin{itemize}
 \item $C^{(3)}$ is symmetric in its three arguments and
       invariant under $\alpha_1\to Q-\alpha_1$.
 \item $C^{(3)}$ vanishes if one of $\alpha_a$
       takes value for degenerate Virasoro representations,
       $\alpha=-mb-nb^{-1}~(m,n\in\mathbb{Z}_{\ge0})$.
 \item $C^{(3)}$ diverges if
       $\alpha_1+\alpha_2+\alpha_3=Q-mb-nb^{-1}~(m,n\in\mathbb{Z}_{\ge0})$,
       since in this case the Liouville
       interaction can screen the violation of momentum conservation.
\end{itemize}
$C^{(3)}_{\alpha_1,\alpha_2,\alpha_3}$ thus has several groups of poles
and zeroes, each group containing infinite number of elements labelled
by two nonnegative integers $m,n$. These should be somehow related to
the eigenvalues of $\susy^2=\epsilon_1J_{12}+\epsilon_2J_{34}+(\cdots)$.

The most interesting would have been the correspondence between
conformal blocks and Nekrasov's partition functions. In the traditional
approach to CFT following \cite{Belavin:1984vu}, the only way to
construct and study conformal blocks was via power series in 2D
coordinate, or in other words summing up all the descendant operators
appearing in the given operator product. There is actually a powerful
recursion relations due to Zamolodchikov \cite{Zamolodchikov:1985ie} that
can determine the coefficients of higher terms in the series expansion
from the lower ones, and it was used in prooving the AGT conjecture
for some basic examples \cite{Fateev:2009aw,Hadasz:2010xp}. A better
understanding of conformal blocks beyond their definition as power
series was definitely needed.  This was a rather unexplored
subject, although Liouville theory has a long history and has played
such an important role in many places in string theory.

Liouville conformal blocks were studied from a different perspective
in a series of work by Teschner
\cite{Teschner:2002vx,Teschner:2003em,Teschner:2003at,Teschner:2005bz}.
As we have seen, conformal blocks form a complete basis of solutions to
conformal Ward identity in a given channel. One can therefore study the
conformal blocks through their transformation property under changes
of basis: namely how the bases of conformal blocks in different channels
are related. Under AGT relation, different channel descriptions of the
same correlator are in correspondence with different Lagrangian
descriptions of the same 4D quantum field theory, that is the
S-duality. On the other hand, it was known that the Liouville conformal
blocks obey the same transformation rule under the change of basis as
the wave functions in quantum Teichm\"uller theory, which is also related to
quantization of the moduli space of flat $SL(2,\mathbb{R})$ gauge fields
on punctured Riemann surface. In \cite{Nekrasov:2010ka,Vartanov:2013ima}
this fact was used as a key to explain how the 4D gauge theories and
Liouville theory are related.

\begin{figure}[ht]
\begin{center}
\includegraphics[width=9cm]{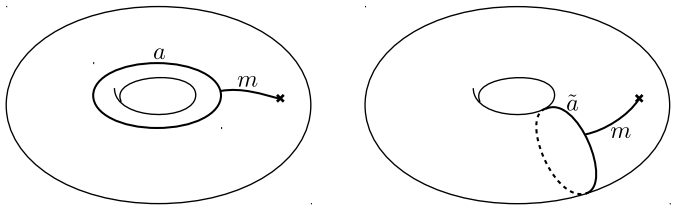}
\end{center}
\caption{Two channels for torus one-point conformal blocks}
\label{fig:sdual}
\end{figure}

For later use, let us look at an example of basis-change of Liouville
conformal blocks for one-point function on the torus. The corresponding
Moore-Seiberg graph is a tadpole, and the conformal blocks are functions
of the modulus $\tau$ of the torus as well the external and internal momenta
$\alpha\equiv\frac Q2+im,\beta\equiv\frac Q2+ia$. See Figure
\ref{fig:sdual}. They transform under modular S-transformation
$\tau\to-1/\tau$ as follows,
\begin{equation}
 {\cal F}_{m,a}(\tau)=\int d\tilde a
 \sinh(2\pi b\tilde a)
 \sinh(2\pi\tilde a/b)\cdot S(a,\tilde a,m){\cal F}_{m,\tilde
 a}(-1/\tau).
\end{equation}
Here we chose a different normalization of conformal blocks
compared to (\ref{eq:corr}). The integral kernel $S(a,\tilde a,m)$
is known to take the following form \cite{Teschner:2003at},
\begin{eqnarray}
 S(a,\tilde a,m) &=& 2^{\frac32}s_b(-m)\int_\mathbb{R}d\sigma\,
  s_b(\sigma+\tilde a+\tfrac m2+\tfrac{iQ}4)
  s_b(-\sigma+\tilde a+\tfrac m2+\tfrac{iQ}4)
\nonumber \\ && \hskip26mm\cdot
  s_b(\sigma-\tilde a+\tfrac m2+\tfrac{iQ}4)
  s_b(-\sigma-\tilde a+\tfrac m2+\tfrac{iQ}4)\cdot e^{4\pi ia\sigma}\,,
\label{eq:S-kernel}
\end{eqnarray}
where $s_b(x)$ is the double-sine function
\begin{equation}
 s_b(x) = \prod_{m,n\in\mathbb{Z}_{\ge0}}
 \frac{\frac Q2+mb+nb^{-1}-ix}{\frac Q2+mb+nb^{-1}+ix}\,.
\end{equation}

\section{3D ${\cal N}=2$ gauge theories}

The idea of SUSY localziation was applied to 3D supersymmetric
Chern-Simons matter systems by Kapustin, Willett and Yaakov (KWY)
\cite{Kapustin:2009kz}. Chern-Simons matter theories are a canonical
example of 3D SCFTs, and some of them are known to have interpretations
as theories of multiple M2-branes. Indeed, the original motivation of
KWY was to provide a precise check of AdS/CFT through the explicit
evaluation of Wilson loops. Moreover, their formula was also applied and
gave an elegant solution to the long standing problem about the growth
$\sim N^{3/2}$ of the degree of freedom on multiple M2-branes. Their
result also found applications and generalizations in many other
interesting problems, some of which we review in the following.

\paragraph{$S^3$ partition function}

KWY constructed supersymmetric Chern-Simons matter theories on $S^3$ and
obtained closed formula for SUSY partition function as well as Wilson
loop expectation values, which apply to a class of 3D ${\cal N}=2$
supersymmetric systems. The system consists of two kinds of multiplets:
a vector multiplet consists of a gauge field $A_m$, gauginos
$\lambda_\alpha,\bar\lambda_\alpha$, real scalar $\sigma$ and an
auxiliary scalar $D$. Chiral multiplet consists of a complex scalar
$\phi$, fermion $\psi$ and a complex auxiliary scalar $F$, and can
couple to vector multiplet in arbitrary representation $R$ of the gauge
group. The gauge fields have Chern-Simons kinetic term
\begin{equation}
 S = \frac k{4\pi}\int\text{Tr}\left(AdA+\frac23A^3\right),
\end{equation}
where $k$ is the quantized Chern-Simons coupling. For each $U(1)$ factor
of the gauge group one can also turn on the Fayet-Iliopoulos coupling
$\zeta$. For chiral multiplets,
in addition to standard gauge interactions one can turn on other
interactions through superpotential, or turn on the so-called real mass through
gauging global symmetry. The supersymmetric Lagrangian and
transformation rules can be written down based on the existence of
conformal Killing spinors on $S^3$,
\begin{equation}
 D_\mu\epsilon ~\equiv~
 \left(\partial_\mu + \frac14\omega_\mu^{ab}\gamma^{ab}\right)\epsilon
 ~=~ \gamma_\mu\tilde\epsilon~~\text{for some }\tilde\epsilon\,.
\label{eq:KSS3}
\end{equation}
An important restriction, to which we will come back later, is that all
the chilal multiplets here are assigned canonical R-charge 1/2. 

The exact $S^3$ partition function depends on $G_k$ (convenient notation
for the gauge group and its Chern-Simons coupling), and chiral matter
representation $R$. The formula reads
\begin{equation}
 Z~=~ \int d^r\sigma e^{i\pi k\text{Tr}(\sigma^2)}
 \prod_{\alpha\in\Delta_+}(2\sinh\pi\alpha\cdot\sigma)^2
 \prod_{w\in R}F(w\cdot\sigma)\,,
\label{eq:ZS3}
\end{equation}
where
\begin{equation}
 F(x)\equiv \prod_{n\ge1}
 \bigg(\frac{n+\frac12+ix}
           {n-\frac12-ix}\bigg)^n = s_{b=1}(\tfrac i2-x)\,.
\label{eq:D-ch}
\end{equation}
The FI coupling $\zeta$ shows up as a
modification of the integrand by $e^{4\pi i\zeta\sigma}$
\cite{Kapustin:2010xq}.

With SUSY localization, the path integral can be shown to simplify to a
finite-dimensional integral over constant values of vector multiplet
scalar $\sigma$, which one can further restrict to the Cartan
subalgebra. An important simplification compared to the four-dimensional
case is the absence of saddle points with non-trivial topological
quantum numbers such as instantons. Another simplification is that the
one-loop determinant here can be evaluated explicitly as a product of
eigenvalues using spherical harmonics, and the evaluation essentially
boils down to representation theory of $SU(2)$. Their formula is thus
very easy to reproduce, so in a sense the 3D theories on $S^3$ can be
thought of as an ideal exercise to learn the essence of SUSY localization.

\paragraph{Application to M2-brane theories}

An important application of KWY formula is to the multiple M2-brane
dynamics and $AdS_4$/CFT$_3$ correspondence. In this area, a long standing
problem was how to understand the growth of the degrees of freedom (or
free energy) on $N$ coincident M2-branes $\sim N^{3/2}$ predicted by
dual supergravity description. If the worldvolume theory on a stack of
$N$ M2-branes is described by a 3D gauge theory with $N\times N$ matrix
valued fields, then the naive count of the degrees of freedom would be
$\sim N^2$. The description of multiple M2-branes worldvolume theory
itself was a long standing problem, but in \cite{Aharony:2008ug} a
${\cal N}=6$ superconformal $U(N)_k\times U(N)_{-k}$ Chern-Simons theory
with bi-fundamental matters was proposed for $N$ M2-branes on orbifold
$\mathbb{R}^8/\mathbb{Z}_k$. Indeed, it is a theory of $N\times N$
matrix valued fields, while the dual supergravity predicts the large $N$
behavior for the free energy
\begin{equation}
 F \sim \frac{\sqrt2\pi}3k^{1/2}N^{3/2}\,.
\label{eq:F-ABJM}
\end{equation}

An elegant solution for this mismatch was proposed in \cite{Drukker:2010nc} by applying
the traditional methods of large-$N$ matrix integrals to the $S^3$
partition function of ABJM model. They in particular found that the
standard 't Hooft expansion of the logarithm of sphere partition
function reproduces (\ref{eq:F-ABJM}) in its leading order. The
subleading contributions as well as instanton contributions were studied
in detail using various approaches to evaluate the integral (\ref{eq:ZS3}),
and interpreted in the dual picture. See the review
\cite{Moriyama:PTEP} for more detail. Note that the fact that the $S^3$
partition function admits such an expansion or resummation is important
in view of AdS/CFT correspondence. The observables in the gauge
theory side need to have well defined analytic continuation in $N$,
because $N$ is mapped to cosmological constant in the gravity side.

\paragraph{AGT relation in 3D}

In 4D gauge theories, one can introduce various defects and study
them. According to their dimensionality they are called loops, surface
defects, or domain walls (or boundaries). It is especially interesting
to study how to describe them using lower dimensional field theories, or
how the duality in 4D gauge theories act on them. Certain domain walls
in 4D ${\cal N}=2$ supersymmetric gauge theories are described by 3D
${\cal N}=2$ field theories, and the sphere partition function gives an
important information on their property.

\begin{figure}[ht]
\begin{center}
\includegraphics[width=7cm]{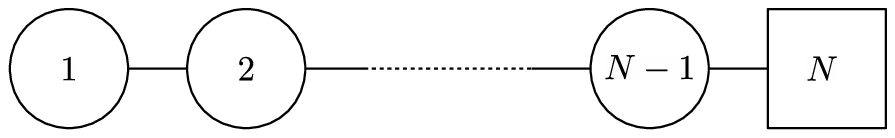}
\end{center}
\caption{Quiver diagram for the theory $T[SU(N)]$}
\label{fig:quiver}
\end{figure}

The study of domain walls and boundaries for this purpose was started in
${\cal N}=4$ SYM by Gaiotto and Witten
\cite{Gaiotto:2008sa,Gaiotto:2008ak}. They were particularly interested
in how the Montonen-Olive $SL(2,\mathbb{Z})$ duality of the SYM acts on
the boundaries and domain walls. As an example, consider the SYM with
gauge group $G$ and take a half-BPS completion of Dirichlet boundary
condition on gauge field. Its S-dual was then shown to be a 3D
${\cal N}=4$ SCFT called $T[G]$ on the boundary coupled to the bulk SYM
with the S-dual gauge group $^LG$. The theory $T[G]$ is
characterized by its global symmetry $G\times G^L$ where $G^L$ is the
gauge group for the S-dual theory. For $G=SU(N)$, the wall
theory has the 3D ${\cal N}=4$ quiver description as in Figure \ref{fig:quiver}.

\noindent
For example, $T[SU(2)]$ is the $U(1)$ SQED with two charged hypermultiplets.
A copy of $SU(2)$ acts as flavor rotation, while another $SU(2)$
isometry shows up as the isometry of the Coulomb branch moduli space
$\mathbb{C}^2/\mathbb{Z}_2$ in the infrared.
The theory $T[G]$ can also be used to describe the S-duality domain
wall, that is the interface where the two ${\cal N}=4$ SYM theory with
gauge groups $G$ and $G^L$ are adjoined.

The structure of S-duality should be even richer for ${\cal N}=2$
supersymmetric theories. As reviewed in the previous section,
two mutually S-dual theories are related in the same way as the
conformal blocks in two different channels are related. Then what kind
of 3D theory shows up at the joint of a pair of mutually S-dual
theories? Though general construction of such theories were not
available, it was conjectured in \cite{Drukker:2010jp} that the $S^3$
partition function of the theory on the wall should correspond to
the transformation coefficients of conformal blocks under changes of
channels, such as the example (\ref{eq:S-kernel}).

An attempt to see the correspondence was made in
\cite{Hosomichi:2010vh}, which studied the S-duality wall between two 4D
half-spaces both supporting ${\cal N}=2^\ast$ theory with $G=SU(2)$. The fields
on the two sides are connected across the wall via S-duality. The vacua
on the two sides are specified by two Coulomb branch
parameters $a,\tilde a$. The theory on the wall was identified as a suitable
mass deformation of the theory $T[SU(2)]$ explained above. In 3D
${\cal N}=2$ terminology, it consists of a $U(1)$ vector multiplet, two
chiral multiplets $q_1,q_2$ of charge $+1$, two chirals
$\tilde q^1,\tilde q^2$ of charge $-1$ and a neutral chiral $\phi$.
The chiral matters all acquire mass proportional to the bulk ${\cal N}=2^\ast$
mass-deformation $m$. In addition, the parameters $a,\tilde a$ enter the
theory as the FI parameter and the mass for charged chirals.

When computing the $S^3$ partition function for the wall theory, a small
but nontrivial problem arose. The 
neutral chiral multiplet $\phi$ of the wall theory is
assigned the R-charge 1, for which the one-loop determinant was not
derived. Without knowing the contribution from $\phi$ it was
proposed in the first version of \cite{Hosomichi:2010vh}
\begin{eqnarray}
 Z_{S^3}(a,\tilde a,m) &=& \text{const}\cdot\int d\sigma\,
 s_{b=1}(\sigma+\tilde a+\tfrac m2+\tfrac i2)
 s_{b=1}(-\sigma+\tilde a+\tfrac m2+\tfrac i2)
 \nonumber \\ && \hskip20mm\cdot
 s_{b=1}(\sigma-\tilde a+\tfrac m2+\tfrac i2)
 s_{b=1}(-\sigma-\tilde a+\tfrac m2+\tfrac i2)\cdot  e^{4\pi i a\sigma}.
\label{eq:ZS3-sdual}
\end{eqnarray}
Though the analysis was incomplete, this result shows quite an agreement with
(\ref{eq:S-kernel}). Thus it was proposed that the AGT relation can be
generalized to include domain walls, and there is a precise relation
between 3D gauge theories and 2D CFTs.

The above observation of the correspondence between 3D gauge theories
and 2D CFTs was soon generalized in an interesting manner. To
explain it, let us recall that the S-duality domain walls are closely
related to Janus domain walls connecting the same 4D gauge theories at
different values of coupling. As a generalization of Janus wall, let us
consider the situation in which the gauge coupling
varies smoothly as a function of one of the spatial coordinates, say $x_3$.
For theories of class S, the situation corresponds to M5-branes wrapping
some Riemann surface whose shape varies as a function of $x_3$. One can
reinterpret it as M5-branes wrapping a 3-manifold. This picture leads to
a correspondence between the geometry of hyperbolic 3-manifolds
${\cal M}$ and the corresponding 3D ${\cal N}=2$ gauge theories
$T[{\cal M}]$, as proposed in \cite{Dimofte:2011ju}.
Moreover, a correspondence which is similar to AGT relation was
proposed between observables of $T[{\cal M}]$ and
Chern-Simons path integrals on ${\cal M}$
\cite{Terashima:2011qi,Dimofte:2011jd,Dimofte:2011ju,Dimofte:2011py},
and various precise correspondence have been reported.

\paragraph{Generalization of KWY formula}

On a closer look at the formulae (\ref{eq:S-kernel}) and
(\ref{eq:ZS3-sdual}), it is tempting to identify $s_b(-m)$ with the one-loop
determinant of the neutral chiral multiplet, as was proposed in the
second version of \cite{Hosomichi:2010vh}.
It is also tempting to look for deformations of the round $S^3$ which
reproduces the double sine function for general $b$, as we will
discuss in the next section.

Finding out the one-loop determinant arising from $\phi$ of
non-canonical R-charge assignment requires generalizing the construction
of supersymmetric theories on $S^3$ accordingly. This turned out
possible, and what is intriguing was that the supersymmetry
transformation rule for chiral multiplet $(\phi,\psi,F)$ then depends
explicitly on its R-charge $q$.
\begin{eqnarray}
 \delta\phi &=& \bar\epsilon\psi,\nonumber \\
 \delta\psi &=& i\gamma^\mu\epsilon D_\mu\phi
 +\tfrac{2qi}{3}\gamma^\mu D_\mu\epsilon\phi+\bar\epsilon F, \nonumber \\
 \delta F &=& i\epsilon\gamma^\mu D_\mu\psi
 +\tfrac i3(2q-1)D_\mu\epsilon\gamma^\mu\psi.
\label{eq:U1R}
\end{eqnarray}
Similar R-charge dependence also shows up in the Lagrangian. The SUSY
localization computation of $S^3$ partition function goes through,
and the one-loop determinant for chiral multiplet of R-charge $q$ was
found to be
\begin{equation}
 F_q(x)\equiv \prod_{n\ge1}\bigg(
 \frac{n+1-q+ix}{n-1+q-ix}\bigg)^n ~=~ s_{b=1}(i-iq-x)\,,
\end{equation}
generalizing (\ref{eq:D-ch}). One can check using this formula that the
neutral chiral multiplet of mass $-m$, R-charge 1 gives rise to the
determinant $s_b(-m)$ which completes the agreement. This was reported
by Jafferis \cite{Jafferis:2010un}, and one day later by \cite{Hama:2010av}.

\paragraph{F-theorem}

Thanks to the above generalization, arbitrary ${\cal N}=2$
supersymmetric theories with R-symmetry can now be put on $S^3$
preserving rigid supersymmetry. For theories with abelian global
symmetry, the assignment of R-charges to chiral matters is not unique;
any two consistent assignments , $q_i=R[\phi_i]$ and
$q'_i=R'[\phi_i]$, differ by a linear combination of abelian global
symmetry charges $Q_a[\phi_i]$. Given a reference R-charge $R_0$,
one can parametrize different assignments of R-charges in the following way,
\begin{equation}
 R ~=~ R_0 +\sum_a t_a Q_a.
\end{equation}
The $S^3$ partition function then becomes a function of the parameters $t_a$. 

If the theory flows to a superconformal field theory in the infrared,
then the R-symmetry in the IR limit is uniquely defined as a member
of the superconformal algebra. Jafferis \cite{Jafferis:2010un} made an
interesting proposal that the corresponding value of $t_a$ can be
determined by extremizing the real part of free energy
$F_{S^3}(t)=-\log Z_{S^3}(t)$. This was proved in \cite{Closset:2012vg}
based on a careful study of the structure of couplings between current
supermultiplets of the field theory and the background supergravity
multiplet.

\section{Squashing}

The comparison of the formulae (\ref{eq:S-kernel}) and (\ref{eq:ZS3-sdual})
leads to another natural guess that the $S^3$ partition function should be
deformed in some way to reproduce the quantities in Liouville theory
with $b\ne1$. We encountered the same unsatisfactory situation also in
the comparison of 4D and 2D observables, but the deformation of $S^3$
gauge theories seems easier to find.

It turned out that the rigid supersymmetry can be realized on manifolds
less symmetric than the round sphere, and moreover one can derive exact
formulae for supersymmetric observables on such manifolds. The important
examples are squashed spheres. It was shown that for a suitable
deformation of the $S^3$ the formula (\ref{eq:ZS3}) is modified to
exhibit the expected $b$ dependence.

\paragraph{Ellipsoid partition function}

One way to achieve $b\ne1$ is to deform the round sphere into ellipsoid
\cite{Hama:2011ea},
\begin{equation}
 \frac1{\ell^2}(x_1^2+x_2^2)+\frac1{\tilde\ell^2}(x_3^2+x_4^2)~=~ 1.
\label{eq:S3b}
\end{equation}
and generalize the Killing spinor equation to include
a background $U(1)_\text{R}$ gauge field $V_\mu$,
\begin{eqnarray}
 D_\mu\epsilon &=&
 \left(\partial_\mu -iV_\mu +\frac14\omega^{ab}_\mu\gamma^{ab}\right)\epsilon
 ~=~ \frac{iH}{2}\gamma_\mu\epsilon,
 \nonumber \\
 D_\mu\bar\epsilon &=&
 \left(\partial_\mu +iV_\mu +\frac14\omega^{ab}_\mu\gamma^{ab}\right)\bar\epsilon
 ~=~ \frac{iH}{2}\gamma_\mu\bar\epsilon\,.
\label{eq:CKS3}
\end{eqnarray}
The scalar function $H$ and the gauge field $V_\mu$ are suitably chosen
so that the above equations have solutions. Then the SUSY localization
leads to the following formula for partition function,
\begin{equation}
 Z ~=~ \int d^r\sigma e^{i\pi k\text{Tr}(\sigma^2)}
 \prod_{\alpha\in\Delta_+}
 4\sinh(\pi b\alpha\sigma)
  \sinh(\pi b^{-1}\alpha\sigma)\cdot
 \prod_{w\in R}s_b\big(\tfrac{iQ}2(1-q)-w\sigma\big)\,,
\label{eq:ZS3b}
\end{equation}
which generalizes (\ref{eq:ZS3}).
The Liouville coupling $b$ was shown to be related to the axis-lengths by
$b=(\ell/\tilde\ell)^{1/2}$.

\paragraph{Sketch of derivation}

The idea of ellipsoidal deformation naturally comes about from the
following observation. In Pestun's derivation of $S^4$ partition
function,  one-loop determinants were evaluated be relating it to the
determinant of the bosonic symmetry $\susy^2$ on some reduced space of
wavefunctions. It is reasonable to expect that $\susy^2$ play similar role in
three dimensions as well.
On the round sphere $x_0^2+x_1^2+x_2^2+x_3^2=\ell^2$, the
localization analysis was based on the Killing spinor
$\epsilon,\bar\epsilon$ satisfying
\[
 \susy^2 = i\bar\epsilon\gamma^m\epsilon\partial_m+\cdots  = \frac i\ell
(J_{12}+J_{34})+\cdots,
\]
where $J_{12},J_{34}$ are the generators of rotations of $\mathbb{R}^4$.
This choice of Killing spinor is essentially unique due to the isometry
of $S^3$. Then a natural guess is that, if there were deformations of
the sphere for $b\ne1$, the square of the corresponding SUSY should
be deformed in the following way,
\begin{equation}
 \susy^2 = i(J_{12}+J_{34})+\cdots
 \quad\longrightarrow\quad
 \susy^2 = ib^{-1}J_{12}+ibJ_{34}+\cdots.
\label{eq:Q2b}
\end{equation}
The deformed geometry therefore should be $U(1)\times U(1)$ symmetric,
and ellipsoids (\ref{eq:S3b}) with the identification
$\ell:\tilde\ell=b:b^{-1}$ is an natural guess. However, at this
level the idea is still too crude, because the conformal Killing spinor
equation (\ref{eq:CKS3}) was known to have solutions only on rather
restricted class of manifolds. Indeed, one can try to solve
(\ref{eq:CKS3}) with various $U(1)\times U(1)$ symmetric ansatz for the
metric and see none of such attempts work except for the round sphere.

During the process of trial and error, we got interested in how the
Killing spinor equation on the round $S^3$ would break down by small
deformations of metric while keeping the Killing spinor unchanged.
Since the problem is to find a family of geometry parametrized by $b$,
one can work perturbatively near $b=1$. If the
small deformation to the geometry were suitably chosen, we could fix the
failure of Killing spinor equation somehow by modifying the Killing
spinor accordingly. As the first experiment, the deformation of the
round sphere into what was traditionally called squashed sphere was considered.
\begin{equation}
 ds^2 = \ell^2(\mu^1\mu^1+\mu^2\mu^2+\mu^3\mu^3)~\longrightarrow~
 ds^2 = \ell^2(\mu^1\mu^1+\mu^2\mu^2)+\tilde\ell^2\mu^3\mu^3.
\label{eq:squash1}
\end{equation}
Here $\mu^a=\mu^a_\mu dx^\mu$ is the basis of left-invariant one-forms
of $SU(2)$. For a suitable choice of Killing spinor $\epsilon$ on the
round sphere, the failure after deformation turned out to be
\begin{equation}
 \Big(\partial_\mu + \frac14\omega_\mu^{ab}\gamma^{ab}\Big)\epsilon
 -\frac{i\tilde\ell}{2\ell^2}\gamma_\mu\epsilon ~=~
 \pm i\Big(1-\frac{\tilde\ell^2}{\ell^2}\Big)\mu^3_\mu\epsilon\,.
\end{equation}
The original plan was to modify $\epsilon$ so that the failure term
(RHS) disappears, but the above equation seemed to suggest a much nicer
alternative solution. One can just regard the failure term as a coupling
to a background vector field $V_\mu$ and include it into the covariant
derivative.

It is a tedious but pleasant exercise to check that the construction of
SUSY transformation rule and Lagrangians all goes through, under the
assumption that $\epsilon,\bar\epsilon$ are assigned the $V_\mu$-charge
$\pm1$. In particular, all the fields can be shown to couple to $V_\mu$
according to their R-charge $q$ in (\ref{eq:U1R}), thus $V_\mu$ can
really be identified as the gauge field for $U(1)$ R-symmetry. At this
point, however, we were not sure what kind of framework would naturally
accommodate this external gauge field. The external gauging of
R-symmetry was regarded just as a tool to define supersymmetry on curved
space, in a similar way to topological twisting.

There was no particular reason to consider the deformation to
traditional {\it squashed sphere} (\ref{eq:squash1}), but in this way
one is left with a large isometry unbroken. The spectrum on
this space can therefore be explicitly solved using spherical harmonics.
In some old literature there are even explicit results on related problems 
\cite{Gibbons:1979kq}. After a detailed spectrum analysis, we found that the
eivenfunctions can be written using spherical harmonics in the
same way as for the round sphere, but the degeneracy of eigenvalues
gets partially resolved due to squashing.  We were hoping that this broken
degeneracy would lead to something new. But dissapointingly, the
one-loop determinants stayed essentially the same as those for the
round sphere.

After all, the square of supersymmetry on the traditional squashed
sphere does not show the expected dependence on $b$
(\ref{eq:Q2b}). Also, the eigenmodes turned out to make nontrivial
contribution to the one-loop determinant as multiplets of the unbroken
$SU(2)$ isometry, so that the determinant still has degeneration of
many zeroes and poles. Thus it looked inevitable to break the isometry
further and try seriously the ellipsoid (\ref{eq:S3b}).

\vskip3mm

Coming back from disappointment, it was pleasing to see that the
ellipsoid (\ref{eq:S3b}) also admits charged Killing spinors if a
suitable background $U(1)_\text{R}$ gauge field $V_\mu$ is turned
on. Moreover, this time the bilinear of Killing spinors indeed showed
the expected $b$-dependence (\ref{eq:Q2b}). The only remaining problem
was how to compute one-loop determinants.

On the ellipsoid (\ref{eq:S3b}) there seemed to be no easy
way to solve the full spectrum. On the other hand, it was clear from
previous experiences of determinant computations that most eigenmodes
form bose-fermi pairs and do not make nontrivial contribution. It is
therefore enough to know the spectrum of the remaining ``unpaired
modes''. It seemed difficult to translate our problem completely
mathematically into the computation of an index as in
\cite{Pestun:2007rz}. Instead, in \cite{Hama:2011ea} the problem was
studied in an equivalent and a little more physical approach by asking the
following questions
\begin{itemize}
 \item what are the Laplace and Dirac operators one wish to know the
       eigenvalues of.
 \item what is the map between the Laplace and Dirac eigenmodes for the
       same eigenvalue.
\end{itemize}
The one-loop determinant can then be expressed by collecting the eigenvalues
of those unpaired modes which are sitting in the kernel and cokernel of
the map. It turned out that all the unpaired modes can be easily listed
up as solutions to some simple first-order differential equations. We thus
arrived at an analytic result (\ref{eq:ZS3b}) which shows precisely the
expected dependence on a new parameter $b=(\ell/\tilde\ell)^{1/2}$.

The analysis of one-loop determinants on the ellipsoid was revisited
later and translated into the computation of indices in
\cite{Drukker:2012sr,Hosomichi:2014hja}.
 
\paragraph{Relation to superconformal index}

It was noticed in \cite{Dolan:2011rp,Gadde:2011ia} that the 3D partition
functions for ${\cal N}=2$ theories have structures similar to
superconformal indices for 4D ${\cal N}=1$ theories. See
\cite{Imamura:PTEP} for a review. The superconformal index is an
observable which encodes the spectrum of BPS operators, and is usually
defined as the trace of time evolution operators over Hilbert space with
an additional insertion of $(-1)^F$. Alternatively, one can use path
integral formulation and define it as a partition function on
$S^1\times S^3$ with SUSY-preserving periodic periodicity condition on
fields. The relation between 3D partition functions and 4D indices were
studied from this viewpoint in \cite{Imamura:2011uw}.

One can introduce a one-parameter deformation to the 4D superconformal
index which is similar to the squashing of 3D partition function by
twisting the periodicity of fields along $S^1$ by isometry rotation of
$S^3$. Interestingly, if the 4D theory with this twist is dimensionally
reduced, the resulting 3D theory is actually on the traditional squashed
sphere (\ref{eq:squash1}), somewhat against our previous observation
which led to the ellpsoid partition function. This led Imamura and
Yokoyama to find another supersymmetric deformation of the round $S^3$
by introducing a background vector field \cite{Imamura:2011wg}.

\paragraph{Further generalization and supergravity}

It is natural to ask what other three-manifolds admits rigid
${\cal N}=2$ supersymmetry, and what is the maximim consistent
generalization of the Killing spinor equation. Festuccia and Seiberg
\cite{Festuccia:2011ws} proposed
that the most suitable framework for such a study is off-shell
supergravity. The background fields introduced in (\ref{eq:CKS3}) or in
\cite{Imamura:2011wg} are then most naturally interpreted
as the (auxiliary) fields in the gravity multiplet, and the Killing spinor
equation is identified with the vanishing of local SUSY transformation
of gravitino. Regarding the existence of rigid supersymmetry on curved
space, it was shown that a 3D space admits a Killing spinor if it
has an almost contact metric structure \cite{Klare:2012gn,Closset:2012ru}.
The general theory of how the 3D partition function can depend on moduli
of almost contact metric structure (such as the squashing parameter $b$)
was developed in \cite{Closset:2013vra}. In particular, it was shown
that partition
function on three-manifolds of the topology of $S^3$ cannot depend on
more than one squashing parameters \cite{Alday:2013lba}.

\paragraph{Squashing $S^4$}

After an instructuve detour to three-dimensions, we finally came back to
the problem of finding a deformation of $S^4$ which reproduces Liouville
correlators with $b\ne 1$. A natural answer was proposed in \cite{Hama:2012bg}
based on the 4D ellipsoid geometry
\begin{equation}
 \frac{x_0^2}{r^2}+\frac{x_1^2+x_2^2}{\ell^2}+\frac{x_3^2+x_4^2}{\tilde\ell^2}
 ~=~1,
\label{eq:S4b}
\end{equation}
with some auxiliary fields in 4D ${\cal N}=2$ off-shell supergravity
turned on. Let us now sketch how this result was derived.

The first step was to identify correct generalization of Killing spinor
equation (\ref{eq:CKS4}), and then use it to construct the
transformation rule and Lagrangian. This analysis was started before the
observation of Festuccia and Seiberg \cite{Festuccia:2011ws}, so the
usefulness of supergravity was not recognized yet. So the only idea to
generalize the Killing spinor equation (\ref{eq:CKS4}) was to turn on
R-symmetry gauge fields. Concerning the metric on the 4D manifold that
realizes $b\ne1$, it seemed natural to assume a fibration structure in
which a 3D ellipsoid is fibered over a segment, with the fiber size
shrinking at the two ends. The ellipsoid (\ref{eq:S4b}) is clearly one
of such examples, where $x_0\in[-r,r]$ is the coordinate on the base
segment and a 3D ellipsoid of varying size is fibered over it. 

It was contrary to our optimistic expectations and even surprising
that the ellipsoid does not admit Killing spinors no matter how one
chooses the R-symmetry gauge field. After a more systematic study of the
ellipsoid-fibration geometries, one 4D metric was found to admit Killing
spinors, but it turned out to have rather strange singularity at
the two poles (points at the end of the segment). It seemed somewhat awkward to
discuss the physics of point-like instantons localized on such a
singular point.

The first nontrivial step was made by recalling that near the north pole
the Killing spinor of our interest should represent the SUSY of
topologically twisted theory on Omega background
$\mathbb{R}^4_{\epsilon_1,\epsilon_2}$.
There the chiral part of Killing spinor $\xi_A$
vanishes while the anti-chiral part $\bar\xi_A$ is finite.
By a suitable gauge rotation one may set
$\bar\xi^{\dot\alpha}_A=\text{const}\cdot\delta^{\dot\alpha}_A$ at the
north pole, since in topologically twisted theory one identifies dotted
spinor indices and $SU(2)$ R-symmetry indices. We also need that the
square of the SUSY give rise to a rotation about the origin
generated by the vector field
\begin{equation}
 v^m~\equiv~
 2\,\bar\xi^A\bar\sigma^m\xi_A
 ~=~(-\epsilon_1x_2\,,\,+\epsilon_1x_1\,,\,-\epsilon_2x_4\,,\,+\epsilon_2x_3),
\end{equation}
where $x_i$ are local Cartesian coordinates near the north pole.
This determines the linear dependence of $\xi_A$ on coordinates
\begin{equation}
 \xi_A~=~\frac12v^m\sigma_m\bar\xi_A\,.
\end{equation}
Now let us perform the failure term analysis in a similar way as in 3D
case. On a flat $\mathbb{R}^4$ without background gauge fields, the
Killing spinor $\xi_A,\bar\xi_A$ satisfies
\begin{eqnarray}
 D_m\bar\xi_A=0,\quad
 D_m\xi_A + \frac18v^-_{kl}\sigma_{kl}\cdot\sigma_m\bar\xi_A =
 \sigma_m\cdot\Big(\frac18v^+_{kl}\bar\sigma_{kl}\bar\xi_A\Big)\,.
\end{eqnarray}
Here $v_{kl}=\partial_kv_l=\partial_{[k}v_{l]}$, and the suffix $\pm$
indicates the self-dual or anti-self-dual components of two-forms.
The failure term is in the second equation, the second term in the LHS.
the tensor $v_{kl}^-$ has nonvanishing components
$v_{12}^-=-v_{34}^-=\frac12(\epsilon_2-\epsilon_1)$, and it vanishes near
the north pole if the squashing deformation is turned off,
$\epsilon_1=\epsilon_2=1/\ell$. The above failure term seemed to suggest
a rather unexpected form of generalized Killing spinor equation,
\begin{eqnarray}
 D_m\xi_A+T^{kl}\sigma_{kl}\sigma_m\bar\xi_A
 &=& -i\sigma_m\bar\xi'_A,
 \nonumber \\
 D_m\bar\xi_A+\bar T^{kl}\bar\sigma_{kl}\bar\sigma_m\xi_A
 &=& -i\bar\sigma_m\xi'_A\quad
 \text{for some }\xi'_A,\bar\xi'_A\,,
\label{eq:gKS41}
\end{eqnarray}
which involves an anti-self-dual tensor $T^{kl}$ and a self-dual tensor
$\bar T^{kl}$ auxiliary fields in addition to the R-symmetry gauge fields
in $D_m$.

It was an enjoyable, though tedious, exercise to construct the
transformation rule and Lagrangian based on the above generalized
Killing spinor equation. One complication was that one needs to require
another set of equations on Killing spinor,
\begin{eqnarray}
 \sigma^m\bar\sigma^nD_mD_n\xi_A+4D_lT_{mn}\sigma^{mn}\sigma^l\bar\xi_A
 &=& M\xi_A,
 \nonumber \\
 \bar\sigma^m\sigma^nD_mD_n\bar\xi_A
 +4D_l\bar T_{mn}\bar\sigma^{mn}\bar\sigma^l\xi_A
 &=& M\bar\xi_A,
\label{eq:gKS42}
\end{eqnarray}
with $M$ another auxiliary field. This looked strange, since this kind
of equations involving square of Dirac operator is usually automatically
satisfied under the assumption of the first-order equations
(\ref{eq:gKS41}).

The proposal of Festuccia and Seiberg came out a little later. 
The generalized form (\ref{eq:gKS41}) of Killing spinor equation turned out all
consistent with the off-shell ${\cal N}=2$ supergravity literature
\cite{deWit:1979ug,deWit:1980tn}, and the fields
$T^{kl},\bar T^{kl},M$ were identified with the auxiliary fields in
gravity multiplet. Also, the additional Killing spinor equation
(\ref{eq:gKS42}) was identified with the local SUSY transformation rule
of an auxiliary spin-1/2 fermion in gravity multiplet, thereby
explaining why the two sets of equations (\ref{eq:gKS41}) and
(\ref{eq:gKS42}) are independent.

The toughest part of the analysis was to show that the ellipsoid
(\ref{eq:S4b}) indeed has Killing spinor for a suitable choice of
background auxiliary fields. The strategy of
\cite{Hama:2012bg} was to assume that a suitably chosen Killing spinor
on the round $S^4$ remains a solution to the Killing spinor equation
after squashing. This requirement turns the Killing spinor equation into a set
of algebraic equations on the background supergravity fields. They
looked highly overdetermined, but turned out to have a family of
solutions which depends on three arbitrary functions invariant under
$U(1)\times U(1)$ isometry. See \cite{Hama:2012bg} for the explicit form
of auxiliary fields. Thus the ellipsoid (\ref{eq:S4b}) was finally shown
to admit rigid supersymmetry.

\paragraph{SUSY localization on $S^4_b$}

The SUSY localization analysis on the ellipsoid \cite{Hama:2012bg}
begins by arguing, based on the continuity in the squashing parameter
$b$, that the SUSY saddle points are parametrized by a constant $a$ in
the same way as on the round $S^4$. Strictly speaking this assumption
should be verified. For the case of 3D squashing the saddle-point
analysis was fully carefully performed in \cite{Alday:2013lba}. Anyway,
once this point is settled, the rest of the analysis is a
straightforward application of localization program.

Again, at all the saddle points the gauge field have to vanish on
generic points on the ellipsoid, but it is allowed to have point-like
instanton or anti-instanton configurations at the two poles. Moreover,
the theory approach near the poles the topologically twisted theory on
$\mathbb{R}^4_{\epsilon_1,\epsilon_2}$ with two independent
Omega-deformation parameters
$\epsilon_1=\ell^{-1},\epsilon_2=\tilde\ell^{-1}$.

Let us finally quickly summarize the essence of one-loop determinant
computation and how it can be reduced to the computation of index. The
computation of one-loop determinant involves a Gaussian integral over
all the fluctuation modes at a give saddle point. Generally, one can
choose as path integration variables a set of bosonic fields
$\bf X$, a set of fermionic fields $\boldsymbol\Xi$ and their superpartners
$\susy\bf X$, $\susy\boldsymbol\Xi$. The supersymmetric measure is then
\begin{equation}
 \langle\cdots\rangle = \int [{\cal D}{\bf X}][{\cal D}(\susy{\bf X})]
 [{\cal D}\boldsymbol\Xi][{\cal D}(\susy\boldsymbol\Xi)]\,(\cdots)\,.
\end{equation}
The one-loop determinant is evaluated as an integral with any $\susy$-exact
Gaussian weight $e^{-\susy{\cal V}}$. Let us take
\begin{eqnarray}
 \susy{\cal V} &=&
 \susy\left\{({\bf X},\susy{\bf X})+(\boldsymbol\Xi,\susy\boldsymbol\Xi)\right\}
 \nonumber \\
 &=& (\susy{\bf X},\susy{\bf X})+(\susy\boldsymbol\Xi,\susy\boldsymbol\Xi)
  +({\bf X},\susy^2{\bf X})-(\boldsymbol\Xi,\susy^2\boldsymbol\Xi)
\end{eqnarray}
Then the one-loop determinant is simply the square root of the ratio of
determinants of $\susy^2$,
\begin{equation}
 Z_\text{1-loop} ~=~ \left(
 \frac{\text{Det}_{\boldsymbol\Xi}(\susy^2)}{\text{Det}_{\bf X}(\susy^2)}
 \right)^{1/2} \,.
\label{eq:sdet}
\end{equation}
It is instructive to see how all these work in examples with finite
number of integration variables. In the toy example of the volume of
sphere (\ref{eq:S2eq}), the supersymmetry $\susy=d-\epsilon i_v$ acts on
the local coordinates ${\bf X}=(x,y)$,
${\boldsymbol\Xi}=(\text{empty set})$ near the north pole as
\begin{equation}
 x\stackrel{\susy}\longrightarrow
 dx \stackrel{\susy}\longrightarrow \epsilon y,\quad
 y\stackrel{\susy}\longrightarrow
 dy \stackrel{\susy}\longrightarrow -\epsilon x.
\end{equation}
The above formula can be used to explain the determinant at the north
pole.

Application of this idea to the path integral of supersymmetric field
theories involves renaming of fields. For example, 4D ${\cal N}=2$
vector multiplet consists of 10 bosons and 10 fermions after
gauge fixing: the physical fields
$A_m,\phi,\bar\phi,\lambda_A,\bar\lambda_A,D_{AB}$, ghosts $c,\bar c$
and Lautrup-Nakanishi field $B$. We take $\susy$ as a combination of
supersymmetry for a specific choice of Killing spinor $\xi_A,\bar\xi_A$
and BRST symmetry, and reorganize these fields under its action. For
example, gauge field $A_m$ is a member of the set ${\bf X}$ whereas its
superpartner
\begin{equation}
 \Psi_m~\equiv~
 i\xi^A\sigma_m\bar\lambda_A-i\bar\xi^A\bar\sigma_m\lambda_A + D_mc
\end{equation}
is a member of $\susy{\bf X}$. The 10+10 fields are thus divided into
four groups ${\bf X}, \susy{\bf X}, {\boldsymbol\Xi}, \susy{\boldsymbol\Xi}$
each consisting of five fields.

The ratio of determinant (\ref{eq:sdet}) can be further simplified if
$\susy^2$ acting on the fields ${\bf X}$ and ${\boldsymbol\Xi}$ has common
eigenvalues. Especially if there is a differential operator $D$ which
relates the fields ${\bf X}$ to ${\boldsymbol\Xi}$ and commutes with $\susy^2$,
then the ratio of determinants can be computed from the index
\begin{eqnarray}
 \text{Ind}(D)
 &=& \text{Tr}_{\bf X}(e^{-i\susy^2t})
    -\text{Tr}_{\boldsymbol\Xi}(e^{-i\susy^2t}) \nonumber \\
 &=& \text{Tr}_{\text{Ker}D}(e^{-i\susy^2t})
    -\text{Tr}_{\text{Coker}D}(e^{-i\susy^2t}).
\end{eqnarray}
Note that the operator $D$ is in principle arbitrary as long as it
commutes with $\susy^2$, and it does not neccesarily have to be related to
Lagrangian of the field theory.
At this point, a powerful localization theorem in mathematics says the
index can be computed as a sum over contributions from $\susy^2$-fixed
points, so we need the precise form of $D$ only near the poles. The
reason of this localization is that $e^{-i\susy^2t}$ involves a finite rotation
(diffeomorphism). If it acts on coordinates as $x^m\mapsto \tilde x^m$, then
the trace of such operator should involve an integral of delta function,
\begin{equation}
d^4x\delta^4(x-\tilde x) = \text{det}(1-\partial\tilde x/\partial x)^{-1}\,.
\end{equation}
so it localizes onto fixed points. For more details see
\cite{Pestun:2007rz,Hama:2012bg} as well as reviews
\cite{Pestun:2014mja,Hosomichi:JPA}.

The one-loop determinant $Z_\text{1-loop}$ for ${\cal N}=2$ gauge
theories on the ellipsoid was thus shown to take the same form
(\ref{eq:Z1l4d}), with the following $b$-dependent modification of the function
$\Upsilon(x)$,
\begin{equation}
 \Upsilon(x)~=~(\text{const})\cdot\prod_{m,n\ge0}
 (mb+nb^{-1}+x)(mb+nb^{-1}+Q-x)\,.
\end{equation}
This function was indeed used to express Liouville three-point functions
\cite{Dorn:1994xn,Zamolodchikov:1995aa}.

\section{Concluding remarks}

Let us briefly mention on the progress in other dimensions.
In five dimensions, the sphere partition function for
supersymmetric gauge theories was studied in
\cite{Kallen:2012cs,Hosomichi:2012ek,Kallen:2012va,Kim:2012ava,
      Kallen:2012zn,Imamura:2012xg,Imamura:2012bm,Kim:2012qf,Kim:2013nva}.
There the important problem was to see how the sphere partition functions
for maximally supersymmetric SYM are related to the index of 6D (2,0)
superconformal theories, and to read off the large-$N$ scaling of the
degrees of freedom on $N$ coincident M5-branes $\sim N^3$.
In two dimensions, the sphere partition function for ${\cal N}=(2,2)$
gauge theories was studied in
\cite{Benini:2012ui,Doroud:2012xw,Gomis:2012wy}. In particular, for
those which flow to ${\cal N}=(2,2)$ superconformal field theories, it
was shown that the sphere partition function computes directly the
K\"ahler potential for the moduli space of superconformal theories.

Localization techniques have been applied to the evaluation of
many supersymmmetric obsercables. In addition to partition functions,
various non-local observables such as Wilson loop 't Hooft loops,
surface operatrors has neen also studied using this technique. They are
not only playing important roles in understanding mathematical
structures underlying supersymmetric gauge theories, but also help us
understanding better how to define and compute such operators precisely
within path integral formalism.

\section*{Acknowledgment}

The author would like to thank Naofumi Hama, Sungjay Lee and Jaemo Park for
collaboration on the materials reviewed in this article. The author is also
gratedul to the Elementary Particle Physics Group at Tokyo Institute of
Technology for giving him opportunity to review the developments in
Liouville theory.


\bibliographystyle{ptephy}
\bibliography{S15E-001-E1-KazuoHosomichi}
%

\end{document}